# Field Effect Transistors based on Networks of Highly Aligned, Chemically Synthesized N=7 Armchair Graphene Nanoribbons

V. Passi[1], A. Gahoi[2], B. V. Senkovskiy[3], D. Haberer[4], F. R. Fischer[4*], A. Grüneis[3], M. C. Lemme[1,2]

[1]AMO GmbH, Advanced Microelectronic Center Aachen, Otto-Blumenthal-Str. 25, Aachen, Germany

[2]Chair of Electronic Devices, RWTH Aachen University, Otto-Blumenthal-Str. 2, Aachen, Germany

[3]II. Physikalisches Institut, Universität Köln, Zülpicher Str. 77, Köln, Germany

[4]Department of Chemistry, University of California Berkeley, Berkeley, California 94720, United States

Corresponding author email: lemme@amo.de, max.lemme@rwth-aachen.de, Tel: +492418867200

*Abstract*— We report on the experimental demonstration and electrical characterization of 7-armchair graphene nanoribbon (7-AGNR) field effect transistors. The back-gated transistors are fabricated from atomically precise and highly aligned 7-AGNRs, synthesized with a bottom-up approach. The large area transfer process holds the promise of scalable device fabrication with atomically precise nanoribbons. The channels of the FETs are approximately 30 times longer than the average nanoribbon length of 30nm to 40nm. The density of the GNRs is high, so that transport can be assumed well above the percolation threshold. The long channel transistors exhibit a maximum $I_{ON}/I_{OFF}$ current ratio of 87.5.

*Index Terms*— 7-AGNRs, back-gated field effect transistors, bandgap, graphene nanoribbons, mobility.

Graphene nanoribbons (GNRs) have been investigated theoretically as tangible nanosystems for many years [1], [2]. The experimental discovery of graphene in 2004 has also led to interest in GNRs in the electronic device community, because unlike two-dimensional graphene, certain GNRs exhibit electronic band gaps due to quantum confinement of charge carriers. This allows proper switching behavior if GNRs are used as a channel material in field effect transistors (GNR-FETs) [3]. In particular, the bandgap in armchair-edged ribbons is roughly inversely proportional to the nanoribbon width ($E_g \sim 1/W$) [1], [4], [5], so that sub-5 nm ribbons will be required for field effect transistors. The semiconducting nature has been experimentally demonstrated in GNR-FETs made from single, isolated graphene ribbons. These have been obtained by top-down nanolithography [6], [7], or through manual selection of solution-phase derived ribbons [8], [9], and "unzipped" carbon nanotubes [10], [11]. These techniques have proved useful to demonstrate the feasibility of GNR electronics, but they are not suitable for delivering atomic precision at sub-5nm dimensions. This includes top-down nanolithography with its severe line edge roughness (at the atomic scale). Bottom-up synthesis with carefully selected precursors, in contrast, has been shown to yield atomically defined graphene nanoribbons and nanostructures [12]. Since such processes rely on metallic catalysts for growth, the GNRs need to be transferred to suitable substrates in order to use them in electronic devices. However, the devices based on bottom-up atomically precise GNRs reported so far were fabricated from GNRs randomly located on substrates, which allows investigating individual GNRs, but is not a scalable technology. Nevertheless, top-gated field effect transistors based on such randomly located individual armchair graphene nanoribbons of $N = 7$ carbon atoms width (7-AGNRs) with 26nm gate length have been demonstrated with on-to-off drain current ratios ($I_{ON}/I_{OFF}$) of $\sim 10^3$ [13]. GNR-FETs with a gate length of 20 nm, thin high-k gate dielectrics and $I_{ON}/I_{OFF}$ ratios of up to $10^5$ have been demonstrated from $N = 9$ atoms wide ribbons [14]. The higher $I_{ON}/I_{OFF}$ ratio in the wider ribbons appears to contradict theoretical expectations – wider ribbons should have a lower band gap - but this can be explained by the different experimental conditions of this emerging technology. In this work, we present GNR-FETs based on dense, highly aligned arrays of 7-AGNRs, similar to [15]. There is great merit in highly aligned, parallel GNRs that form one transistor (or sensor) with multiple nanoribbon channels: such parallelism will be required to meet the demands of drain currents in the transistor on-state, i.e. a single ribbon will not be able to deliver the required performance. This is very similar to work on carbon nanotubes, where highly aligned CNTs (and their networks) are under intense investigation even after nearly 20 years of the first demonstration of a CNT FET [16]–[20]. Such arrays can potentially be scaled up to allow wafer-scale fabrication of atomically precise GNRs as a material platform for electronic devices. Highly parallel GNR channels have great potential for thin film transistor applications, just like their CNT counterparts [18], [21], [22]. In addition, long channel GNRs would provide ultimate sensitivity for chemical and biosensors [23]–[25].

The 7-AGNRs were synthesized on Au(788) crystal using 10,10-dibromo-9,9-bianthryl (DBBA) molecules and a chemical bottom-up approach [12], [26]. Figure 1a shows the STM topographic image of the obtained system. This growth procedure results in densely aligned 7-AGNRs with an average length of the ribbon of approximately 30 nm. A small distance between the individual parallel ribbons (~1nm) leads to the inter-ribbon interaction and up-shift of the Raman modes in comparison to the system on Au(111) [27]. The synthesized layers of aligned 7-AGNRs were transferred from Au(788) onto an oxidized silicon substrate using an electrochemical delamination technique, preserving the alignment of the nanoribbons [28]. The high quality and orientation of the transferred layer was confirmed by polarized Raman measurements, which are shown in Figure 1b. The Raman spectra of narrow AGNR contain several characteristic peaks, which are not observed for graphene (particularly at 397, 960, 1222 and 1262 cm$^{-1}$). In analogy with carbon nanotubes, the peak at ~397cm$^{-1}$ is called radial-breathing-like-mode (RBLM). The frequency of RBLM strongly depends on the width of armchair ribbon [29] and its energy position strongly indicates that we indeed have a width of only 7 carbon atoms. The peaks at 1608 cm$^{-1}$ and 1344 cm$^{-1}$ are called G- and D-like modes, since they resemble the atomic vibrations in 2D graphene sheet. Note, however, that the D-like mode observed in 7-AGNRs originates from the phonon with momentum $q = 0$, in contrast to graphene (supporting information in [27]).

The transferred GNRs were cleaned with acetone followed by an iso-propanol rinse. Contact patterns with varying spacing were defined by electron beam lithography with a bilayer PMMA resist stack. After resist development, a stack of nickel / gold (20 nm / 250 nm) was deposited and excessive metal was removed by lift-off. A thin layer of aluminum was deposited at the bottom of the substrate to enable back-gate ($V_{bg}$) biasing during electrical measurements. The fabrication process sequence for these back-gated devices is shown in Figure **2**a. A scanning electron micrograph of the fabricated device with unpatterned GNRs is shown in Figure **2**b while the inset shows one complete device including probe pads. The source (S) and drain (D) contacts were deposited perpendicular to the direction of GNR alignment (red lines in Figure **2**b) in order to allow current propagation along the ribbon axis.

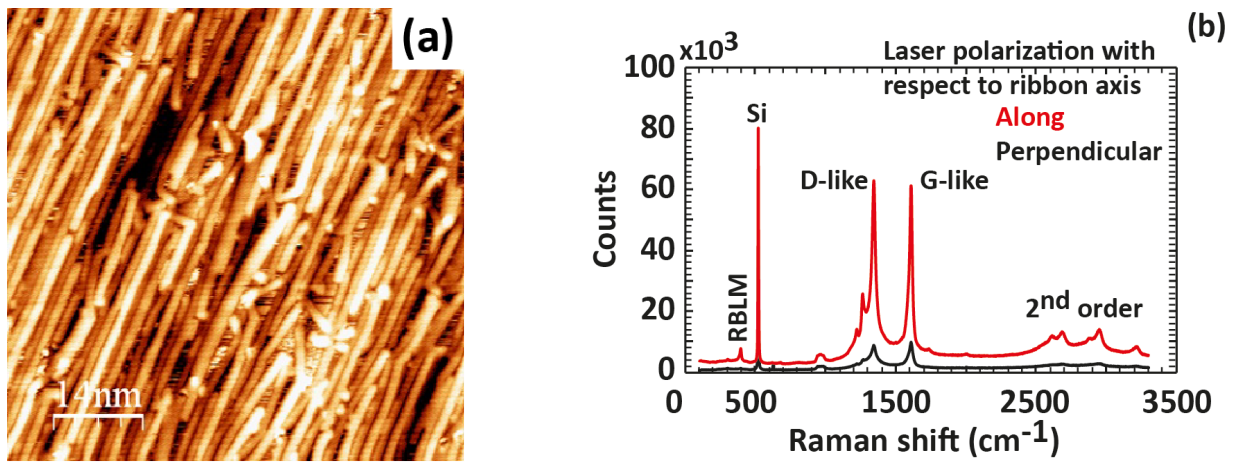

Figure 1: Characterization of the graphene nanoribbons. (a) Scanning tunneling microscope images of the 7-AGNRs taken on Au (788) ($V_s$ =0.6V and $I_t$ = 40pA). (b) Polarized Raman spectrum of the 7-AGNRs on silicon-oxide showing the radial-breathing-like-mode (RBLM), D-like and G-like peaks as well as the second order Raman peaks.

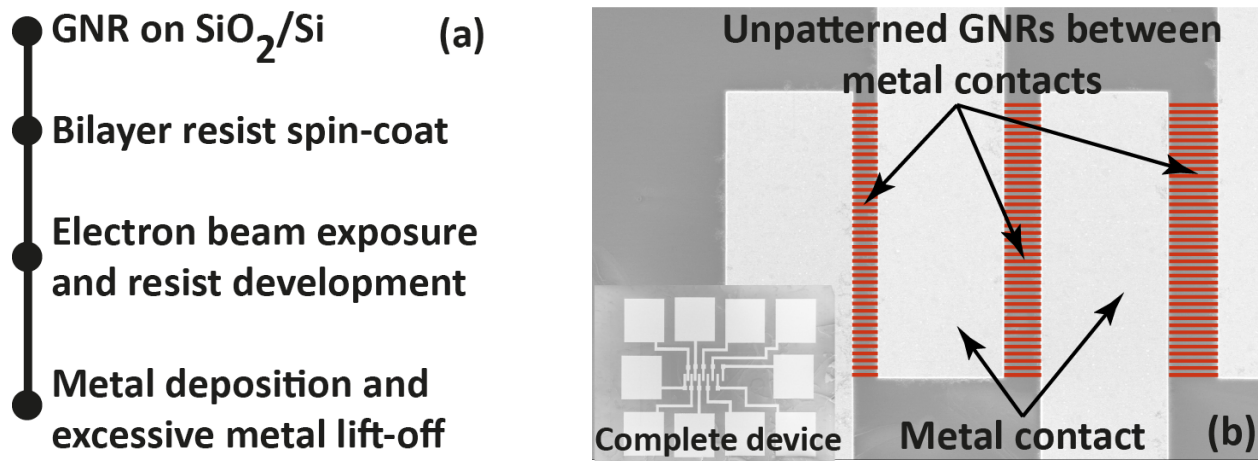

Figure 2: (a) Sequence of process steps for device fabrication. (b) Scanning electron micrograph of a device with metal contacts (with varying spacing) on highly parallel GNRs. The inset shows the entire device including contact pads. The red lines are a guide to the eye and indicate the direction of the densely aligned GNRs.

Electrical measurements were carried out under vacuum ($3.2 \times 10^{-4}$ mbar) at room temperature of 21°C (Figure **2**). The output characteristics (drain current $I_d$ vs. source-drain voltage $V_{ds}$) of the device shown in Figure 3a exhibit clear gate bias ($V_{bg}$) dependence. We further observe asymmetric drain currents ($I_d$) for the hole and electron branch, with lower current levels for positive drain bias and higher current levels for negative drain bias. This asymmetry is attributed to the work function difference of the contact metal (nickel) and the semiconducting GNRs, which lead to asymmetric injection of electrons and holes, similar to effects previously reported in carbon nanotube field effect transistors [30]. The presence of a Schottky barrier at the metal-GNR interface is evident by the non-linear behavior at low drain-source bias ($V_{ds}$). Given the µm-scale S/D distance and the short individual GNR length, the charge transport in the devices can then be interpreted as a combination of diffusive and ballistic transport (through individual GNRs) and transport through a percolation network well above the percolation limit.

Transfer characteristics of the 7-AGNR-FET are shown in Figure 3b for different source-drain bias voltages. The gate leakage current is in the range of $10^{-10}$ A and $10^{-12}$ A, which is two orders of magnitude lower than the measured drain current. A maximum $I_{ON}/I_{OFF}$ ratio of 87.5 is extracted at $V_{DS} = -2$ V. The field effect mobility is 0.00122 cm²/Vs, calculated using the direct transconductance method [31]. We stress that this number is not representative of the individual GNRs, but rather owed to the percolation network. Furthermore, it is an underestimation due to the rather high contact and ribbon-to-ribbon resistances. We expect a drastic reduction in contact resistance with proper contact engineering [32], [33]. In addition, various factors such as surface charge traps, interfacial phonons, substrate-stabilized ripples and potential contamination due to the transfer and fabrication processes may further limit performance. Nevertheless, the high $I_{ON}/I_{OFF}$ ratio (compared to large-area graphene) and the overall switching behavior at room temperature are clearly indicative of an electronic band-gap in the 7-AGNR channels. The present work on long channel devices from networks of nanoscale GNRs is only a first step towards demonstrating the feasibility of the approach. Despite channel lengths that are approximately 30 times larger than the average ribbon length, the devices exhibit semiconducting behavior with a maximum $I_{ON}/I_{OFF}$ current ratio of 87.5. The next logical step will be to move towards GNR-FETs with gate lengths below the length of the individual ribbons. This can be met either by employing nanolithography or by modifying the growth process to obtain longer ribbons. In addition, the density of parallel ribbons may be further defined through growth processes in the future in order to obtain parallel, but not overlapping GNRs. On the device level, substantial performance improvements can be expected by defining channel areas with patterned AGNRs, thin top-gate oxides and optimized source and drain metal contacts. With these further optimization options expected to improve device performance, our work indicates a way towards utilizing massively parallel and aligned GNRs for integrated devices and circuits.

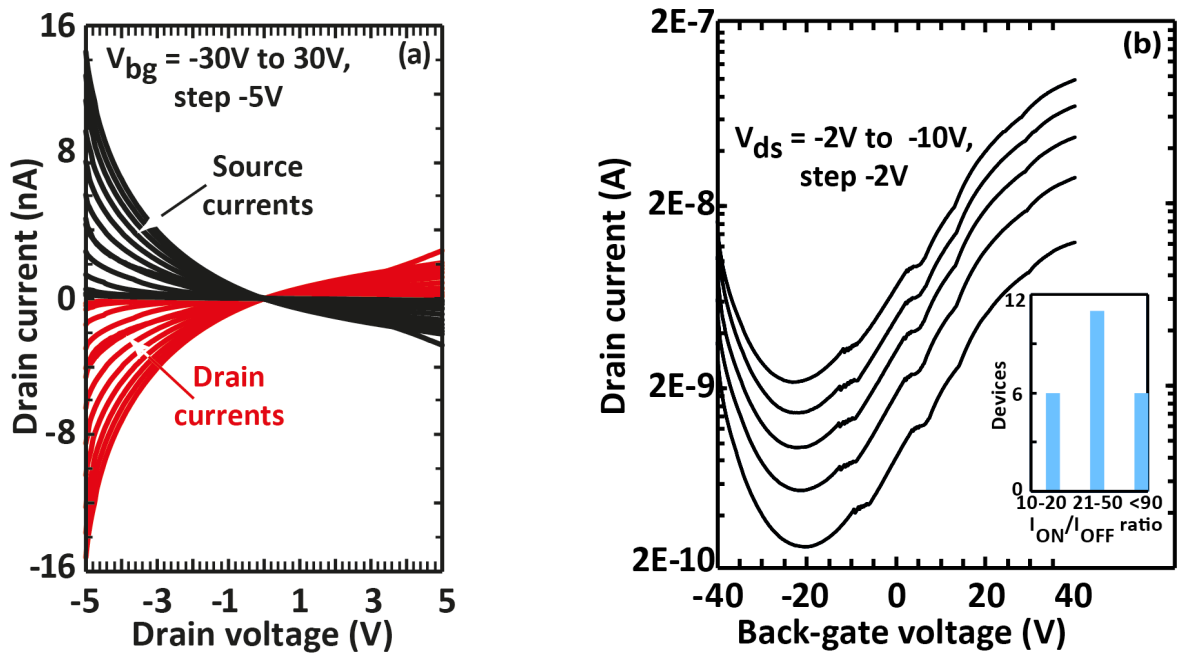

Figure 3: Electrical characterization of a 7-AGNR field effect transistor. (a) Output characteristics at varying back-gate bias and (b) transfer characteristics at varying drain-source bias (inset: histogram of $I_{ON}/I_{OFF}$ current ratios measured in different devices).

**Methods**

The starting substrates consisted of highly dense, aligned 7-AGNRs transferred onto oxidized silicon. The sample was cleaned using warm acetone at 40°C for 10mins, rinsed in IPA and dried with nitrogen. The sample were baked on a hot plate at 180°C for 10 mins followed by PMMA spin-coating. Two layers of PMMA were used: a bottom layer of copolymer EL-13% was spin-coated at 3000 rpm followed by a bake at 180°C for 10 min, and a top layer of PMMA-3% 495k was spin coated at 2500 rpm followed by a bake at 180°C for 10 mins. The total resist thickness was 650 nm, measured by ellipsometry. Electron beam lithography (Raith EBPG 5000Plus) was performed at 100keV using the following parameters: a dose of 420μC/cm² using proximity correction, a current of 20 nA and a resolution of 25 nm. After metal deposition by evaporation, excessive metal was removed by placing the sample in warm acetone at 40°C for 1 hr. Finally, the samples were rinsed with IPA and nitrogen blow dried.

Electrical characterization was performed using a Lakeshore probe station connected to a Keithley SCS4200 semiconductor parameter analyzer. All measurements were carried out under vacuum ($3.2 \times 10^{-4}$ mbar) at a temperature of 21°C. The drain current was measured while sweeping the back-gate bias from –40 V to +40 V in voltage steps of 0.2 V with a hold / delay times of 0.5 s and 0.5 s, respectively.

**Acknowledgements**

The authors acknowledge support through ERC grants 648589 "SUPER-2D" and 307311 "InteGraDe", funding from DFG projects CRC 1238 (project A1), GR 3708/2-1 and LE 2440/2-1, European regional funds grant NW-1-1-036b "HEA2D" and the U.S. Department of Energy, Office of Science, Office of Basic Energy Sciences, under Award no. DE-SC0010409 (design, synthesis, and characterization of molecular precursors).

F.R.F. is also with Materials Science Division, Lawrence Berkeley National Laboratory, Berkeley, California 94720, United States and Kavli Energy Nanosciences Institute at the University of California Berkeley and Lawrence Berkeley National Laboratory, Berkeley, California 94720, United States.